\documentstyle[aps,prl,floats,twocolumn,psfig,epsfig]{revtex}

\def\ub{{\overline{u}}}
\def\vb{{\overline{v}}}

\begin{document}

\twocolumn[\hsize\textwidth\columnwidth\hsize\csname@twocolumnfalse\endcsname
\draft
\title
{Chiral phase transitions: focus driven critical behavior
in systems with planar and vector ordering}

\author{
P. Calabrese${}^{1}$,
P. Parruccini${}^2$, and
A. I. Sokolov${}^{1,3}$\cite{pa}
}

\address{$^1$Scuola Normale Superiore and INFN,
Piazza dei Cavalieri 7, I-56126 Pisa, Italy.}

\address{$^2$Dipartimento di Fisica dell' Universit\`a di
Pisa and INFN, Via Buonarroti 2, I-56100 Pisa, Italy.}

\address{$^3$Department of Physical Electronics,
Saint Petersburg Electrotechnical University,
Professor Popov Street 5, St. Petersburg 197376, Russia.\\
{\bf e-mail: \rm
{\tt calabres@df.unipi.it},
{\tt parrucci@df.unipi.it},
{\tt ais@sokol.usr.etu.spb.ru},
}}

\date{\today}

\maketitle

\csname@twocolumnfalse\endcsname

\begin{abstract}
The fixed point that governs the critical behavior of magnets
described by the $N$-vector chiral model under the physical
values of $N$ ($N =2, 3$) is shown to be a stable focus both
in two and three dimensions. Robust evidence in favor of this
conclusion is obtained within the five-loop and six-loop
renormalization-group analysis in fixed dimension.
The spiral-like approach of the chiral fixed point results in
unusual crossover and near-critical regimes that may imitate
varying critical exponents seen in physical and computer
experiments.
\end{abstract}

]

The critical behavior of the $N$-vector chiral model remains
a controversial issue, being, at the same time, of prime
interest for those studying phase transitions in helical
magnets, stacked triangular antiferromagnets, and frustrated
Josephson-junctions arrays both theoretically and experimentally.
In principle, this model should reproduce the main features
of phase transitions into chiral states in above substances,
but various approaches and approximation schemes still yield
quite different theoretical results (see
\cite{Kawamura-98,cp-97,rev-01} for review and
\cite{lsd-00,tdm-00,prv-01p,prv-01n,prv-02,i-01,cc-01,rj-92,olsson-95,md-01,fckf-00,ss-99} for most recent
three- and two-dimensional results).
Even within the same, otherwise
powerful field-theoretical renormalization-group (RG) approach
the theoretical predictions are found to strongly, on the
qualitative level, depend on the order of approximation
and the reference point of the $\epsilon$-expansion
\cite{prv-01p,cc-01,Kawamura-88,adj-90,as-94,asv-95}.
The experimental data for the critical exponents obtained in
several relevant materials turn out also to be considerably
scattered (see
\cite{Kawamura-98,cp-97,rev-01,prv-01p,pkvmw-00,psbkgrv-01,fdp-02}
and references therein).

In this work, we put forward the idea that the controversial
situation one faces studying the phase transitions in the
aforementioned systems may reflect the quite unusual mode of
critical behavior appropriate to the $N$-vector chiral model
under the physical values of $N$. To argue this conjecture, we
analyze the critical thermodynamics of the two-dimensional
and three-dimensional chiral models within the five-loop and
six-loop renormalization-group approximations in fixed
dimension and determine the
structure of RG flows for $N =2, 3$. In the course of this
study, the advanced resummation technique is used that proved
to be effective even near the borderline where the divergent RG
series loose their Borel summability \cite{prv-01p,cpv-00,pv-00}.
We will show that the critical behavior of the systems with
planar and vector chiral ordering is governed by the stable
fixed point which is a focus attracting RG trajectories in a
spiral-like manner. Approaching the fixed point in a non-monotonic
way, looking somewhat irregular, may result in a large variety of
crossover and near-critical regimes that give rise to strongly
scattered effective critical exponents seen in numerous
experiments and computer simulations.

We start with the Landau-Ginzburg-Wilson Hamiltonian of the
$D$-dimensional chiral model
\begin{eqnarray}
{\cal H} = \int d^D x
&& \left\{ {1\over2}
\sum_{a} \left[ (\partial_\mu \phi_{a})^2 + r \phi_{a}^2 \right]
+ {1\over 4!}u_0 \left( \sum_a \phi_a^2\right)^2 \right. \nonumber\\
&& \left. + {1\over 4!} v_0
\sum_{a,b} \left[ ( \phi_a \cdot \phi_b)^2 - \phi_a^2\phi_b^2\right]
 \right\},\label{LGWH}
\end{eqnarray}
where $\phi_a$, $a = 1, 2$ are two $N$-component vectors.
Under $v_0 > 0$ this Hamiltonian describes transitions
into the phase with noncollinear ordering, i.e. into the chiral
state. Since the lower-order renormalization-group approximations
failed to give consistent results, we are in a position to study
the critical behavior of the model (1) on the base of RG series
of maximal length. For the three-dimensional chiral model such
record, six-loop series were calculated in \cite{prv-01p}, while
in two dimensions only four-loop RG expansions were known
\cite{cc-01}. Aiming to obtain more reliable results, we extend
the latter series to the five-loop order, using numerical
values of the five-loop integrals evaluated in \cite{os-00};
explicit expressions for the five-loop contributions to
$\beta$-functions and critical exponents will be published
elsewhere \cite{cops}.

The renormalization-group expansions are known to be divergent
and some resummation procedure should be applied to extract
proper physical predictions and numerical estimates.
Taking into account the properties of Borel summability and the
large order behavior of the series in two \cite{cc-01} and three \cite{prv-01p}
dimensions,
we resum the perturbative expressions by means of a Borel
transformation combined with an appropriate conformal
mapping \cite{LZ-77} for the analytical extension of the
Borel transform.
With this resummation procedure we obtain the following approximants
for each perturbative series $R(\ub,\vb)$:
\begin{eqnarray}
E({R})_p(\alpha,b;\bar{u},\bar{v})&=& \sum_{k=0}^p
  B_k(\alpha,b;\bar{v}/ \bar{u}) \nonumber \\
 & \times & \int_0^\infty dt\,t^b e^{-t}
  {y(\bar{u} t;\bar{v}/\bar{u})^k\over [1 - y(\bar{u} t;\bar{v}/ \bar{u})]^\alpha}
\label{approx}
\end{eqnarray}
where
\begin{equation}
y(x;z) = {\sqrt{1 - x/\overline{u}_b(z)} - 1\over
 \sqrt{1 - x/\overline{u}_b(z)} + 1},
\end{equation}
and $\ub_b$ is the singularity of the Borel transform
(evaluated in Ref. \cite{prv-01p,cc-01})
closest to the origin at fixed $z=\vb/\ub$.
Varying two parameters $\alpha$ and $b$ we are able to estimate
the systematic error of the final results.
We consider the 18 approximants for each $\beta$ function that minimize
the difference between perturbative orders:
the approximants with $b=5,7,\dots ,15$ and $\alpha =0,1,2$
are used for the two-dimensional models \cite{cc-01},
whereas in three dimensions we choose the approximants with
$b=3,6\dots,18$ and $\alpha=0,2,4$ \cite{prv-01p}.

Applying this procedure to the evaluation of the common
zeros of the $\beta$ functions, we find in three
dimensions a chiral fixed point, as in \cite{prv-01p}.
In two dimensions the five-loop results 
also confirm the existence of the chiral fixed
point found earlier at the four-loop level \cite{cc-01}.

Then we look at the stability properties of these fixed
points. Since the exponents $\omega_{+}$ and $\omega_{-}$
determining the fixed point stability are eigenvalues of
the matrix
\begin{equation}
\Omega = \left(\matrix{\displaystyle \frac{\partial \beta_\ub(\ub,\vb)}{\partial \ub}
 &\displaystyle \frac{\partial \beta_\ub(\ub,\vb)}{\partial \vb}
\cr &
 \cr \displaystyle \frac{\partial \beta_\vb(\ub,\vb)}{\partial \ub}
& \displaystyle \frac{\partial \beta_\vb(\ub,\vb)}{\partial \vb}}\right)\; .
\label{stability-matrix}
\end{equation}
we consider the numerical derivatives of each pair of
approximants of the two $\beta$-functions at their common zero.
The majority of the eigenvalues, evaluated in two and three
dimensions using 324 combinations of the approximants for each case,
has non-vanishing imaginary parts. Namely, for $D=2$, $N=2$ the
complex eigenvalues are produced by 87\% of working
approximants, for $D=2$, $N=3$ - by 89\%, and for $D=3$,
$N=2$ - by 76.5\%. For the physically important
three-dimensional model with $N=3$, all the approximants
employed are found to give the eigenvalues with non-zero
imaginary parts. To have reliable numerical estimates
for $\omega_{+}$ and $\omega_{-}$, we limit ourselves by the
approximants that yield the chiral fixed point coordinates
compatible with their final, properly weighted values. The
approximates leading to purely real eigenvalues of the
stability matrix are discarded in the course of the averaging
procedure.

\begin{figure}[t]
\centerline{\psfig{height=6truecm,width=8.6truecm,file=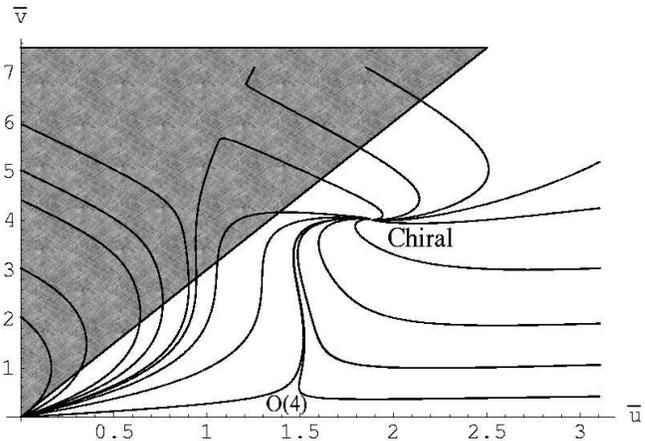}}
\caption{RG flow for the physically important case $D=3$, $N=2$.
Both $\beta$-functions are represented by the approximants with
$\alpha=2$ and $b=6$. With this choice, the chiral fixed point
coordinates $(\ub, \vb)$ are $(1.882,4.017)$.}
\label{fign2d3}
\end{figure}

Thus, for the three-dimensional model with $N=2$ we obtain
\begin{eqnarray}
\omega_{\pm}=&1.40(35)\pm i \, 1.00(45) ,& \quad 5\;loop, \\
\omega_{\pm}=&1.70(40)\pm i \, 0.90(40) ,& \quad 6\;loop,
\end{eqnarray}
while for $N=3$
\begin{eqnarray}
\omega_{\pm}=&0.85(20)\pm i \,0.95(15) ,& \quad 5\;loop, \\
\omega_{\pm}=&1.00(20)\pm i \,0.80(25) ,& \quad 6\;loop,
\end{eqnarray}
In two dimensions we have for $N=2$
\begin{eqnarray}
\omega_{\pm}=& 1.50(25)\pm i \,1.00(45), & \quad 4\;loop, \\
\omega_{\pm}=& 2.05(35)\pm i \,0.80(55), & \quad 5\;loop,
\end{eqnarray}
and for $N=3$
\begin{eqnarray}
\omega_{\pm}=& 1.30(25)\pm i \,0.50(35) , & \quad 4\;loop, \\
\omega_{\pm}=& 1.55(25)\pm i \,0.55(35) ,& \quad 5\;loop.
\end{eqnarray}
The appearance of imaginary parts comparable in magnitude
with the real ones clearly indicates that the chiral critical
behavior is controlled by the focus fixed point both in three
and two dimensions \cite{noteNA}.

To further confirm this conclusion, we apply also 
the resummation by means of the
Pad\'e-Borel-Leroy technique (see e.g. Refs. \cite{rev-01,os-00}). 
The results are substantially
equivalent, although the statistics is much less significant
due to a large number of defective Pad\'e approximants 
(i.e. possessing dangerous poles)  generated during the resummation.

\begin{figure}[t]
\centerline{\psfig{height=6truecm,width=8.6truecm,file=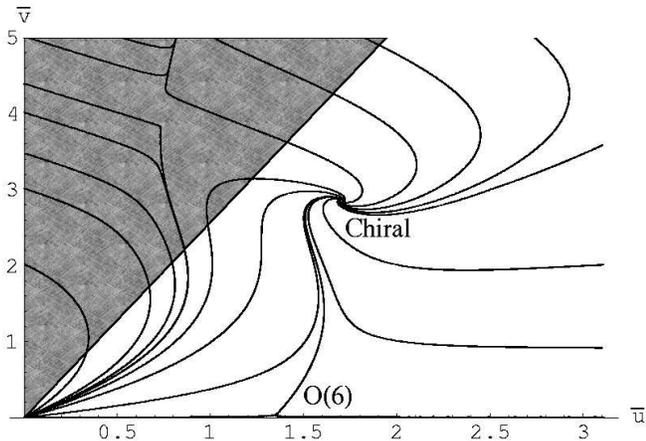}}
\caption{RG flow for the physically interesting case $D=3$, $N=3$.
Both $\beta$-functions are given by approximants with $\alpha=2$ and
$b=9$, the chiral fixed point is located at $(1.702,2.858)$.}
\label{fign3d3}
\end{figure}

\begin{figure}[t]
\centerline{\psfig{height=6truecm,width=8.6truecm,file=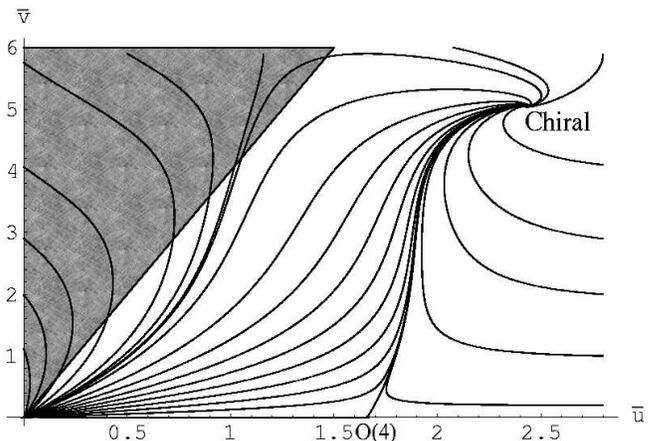}}
\caption{RG flow for the two-dimensional model with $N=2$.
The approximants used are given by $\alpha=1$ and $b=5$
for $\beta_\ub$ and $\alpha=0$ and $b=8$ for $\beta_\vb$.
In this case, the chiral fixed point is at $(2.427,5.045)$.}
\label{fign2d2}
\end{figure}

\begin{figure}[b]
\centerline{\psfig{height=6truecm,width=8.6truecm,file=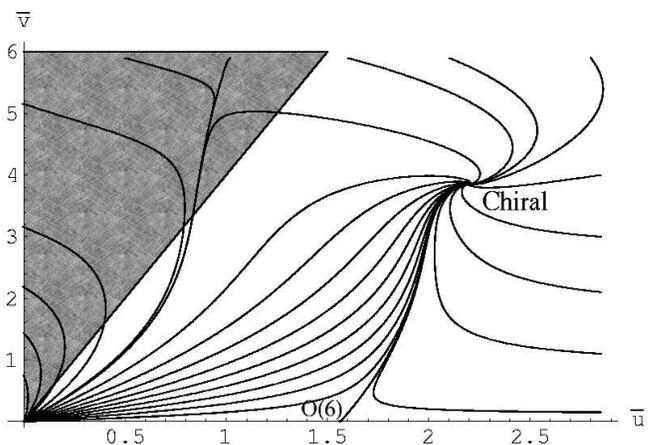}}
\caption{RG flow for $D=2$, $N=3$.
The approximants used are given by $\alpha=1$ and $b=5$
for $\beta_\ub$ and $\alpha=0$ and $b=8$ for $\beta_\vb$,
the chiral fixed point coordinates are $(2.201,3.854)$.}
\label{fign3d2}
\end{figure}

In order to substantiate how the focus driven critical
behavior manifests itself in the RG flows, we investigated
the structure of these flows, given by the resummed five-loop
($2D$) and six-loop ($3D$) RG expansions. Here, we
report four examples of RG flows corresponding to physical
values of $N$ and generated, for certain values of $D$,
by the typical working approximants.
They are presented in Figs. \ref{fign2d3}, \ref{fign3d3},
\ref{fign2d2}, and \ref{fign3d2}, clearly demonstrating
the spiral-like approach of the chiral fixed point.
Obviously, all the RG flows are quantitatively correct
only within the regions where the singularity of the Borel
transform closest to the origin is on the real negative
axis (see \cite{prv-01p,cc-01}); in Figs. \ref{fign2d3},
\ref{fign3d3}, \ref{fign2d2}, and \ref{fign3d2} they
correspond to unshaded areas. Nevertheless, we report
also the flows in other parts of the coupling constant
plane to present a complete qualitative picture \cite{note1}.

From these figures another peculiar and interesting feature
emerges. For some manifold of the starting points, the RG
trajectories have the coordinate $v$ that grows very fast
at the beginning and seem to reach the region of the first
order phase transitions, but just before arriving there these
trajectories drastically curve moving toward the stable chiral
fixed point. Such RG flows may be considered as a possible
explanation of recent Monte Carlo simulations \cite{i-01} in
which a similar behavior was observed (in the space of Binder
cumulants), but only up to the first rising up of the coupling
$v$, interpreted as an evidence of fluctuation induced first
order transition.

Within the context of the results obtained, it is reasonable
to discuss the fortune of the unstable {\it anti-chiral} fixed
point. For almost all the working approximants, this point is
seen in the region where the singularity of the Borel
transform closest to the origin is on the real positive axis,
leaving the existence of the anti-chiral point doubtful. The fact
that its location strongly oscillates with varying the
approximants indicates that in this domain of $(u, v)$ plane
the analysis is not robust. On the other hand, under the
presence of the stable focus chiral point, the second, unstable
fixed point is not topologically needed on the separatrix
dividing the regions of first and second order phase
transitions.

\begin{figure}[t]
\centerline{\psfig{height=3.9truecm,width=8.66truecm,file=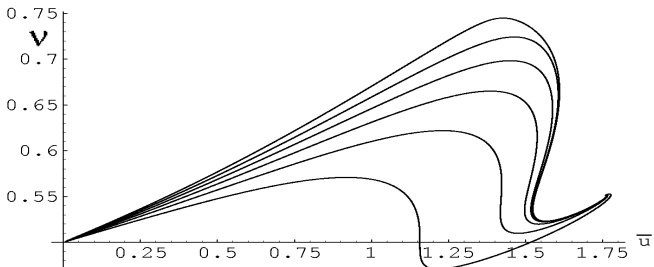}}
\caption{Effective exponent $\nu$ for $N=3$ and $d=3$, along RG trajectories.
Note that $\nu$ is a multi-valued function of $\ub$, as a consequence of the
non-monotonic behavior of the RG trajectories.}
\label{cross}
\end{figure}

Finally we report in Fig. \ref{cross} the effective exponent $\nu$ for $N=3$ 
and $d=3$, evaluated along the RG trajectories of Fig. \ref{fign3d3}.
The exponent oscillates within a large range ($0.47\leq \nu\leq 0.72$), even
close to the stable fixed point. This provides a possible explanation of the 
scattered results obtained in simulations and experiments; 
in fact the majority of numerical results, obtained when a continuous phase
transition was observed, are in this wide range.

It is worthy to note that earlier the focus-like stable fixed
points were found on the RG flow diagrams of the model
describing critical behavior of liquid crystals \cite{ksh-79}
and of the $O(n)$-symmetric systems undergoing first order phase
transition close to the tricritical point \cite{s-79}.
In these cases, however, the independent coupling constants
had different scaling dimensionalities and played essentially
different roles in forming the critical thermodynamics.
Moreover, recently the focus driven chiral phase transition
was observed in the three-dimensional model (1) within the
three-loop approximation, but for quite unphysical values of
$N$ ($N = 5, 6, 7$) \cite{lsd-00}.
For real physical systems having coupling constants
of the same scaling dimensionality, the robust evidence
in favor of phase transitions governed by the focus
stable fixed point is presented for the first time.

We thank E. Vicari for discussions of the obtained
results. The financial support of the Russian Foundation
for Basic Research under Grant No. 01-02-17048 (A.I.S.)
and the Ministry of Education of Russian Federation under
Grant No. E00-3.2-132 (A.I.S.) is gratefully acknowledged.
A.I.S. has benefitted from the warm hospitality of Scuola
Normale Superiore and Dipartimento di Fisica
dell'Universit\`a di Pisa, where this research was done.

\begin{references}
\bibitem{note1}
We adopt here different rescalings for the four-point
renormalized couplings in two and three dimensions; they
coincide with those used in Refs. \cite{cc-01} and
\cite{prv-01p}, respectively. Due to this difference, the
shaded areas turns out to depend on the space dimensionality $D$.
In these pictures the mean-field domain of first-order phase
transitions are exactly
two times larger than the shaded areas~($z>2$ in the symmetric
normalization of Ref. \cite{cc-01}).
Note that saying "the domain of the first-order phase
transitions" we mean, as usually, the region where the
quartic form in free energy expansion can acquire negative
values. In fact, because of the presence of the higher-order
terms in this expansion that make the system globally stable
at any temperature, the true domains of the first-order
transitions may be substantially more narrow than those
predicted by the mean-field approximation.
\end{references}
\end{document}